\documentclass[a4paper,11pt]{article}
\usepackage{pos}

\title{Directed flow induced by electromagnetic fields in heavy ion collisions}

\author*[a]{Yifeng Sun}
\author[b,a]{Salvatore Plumari}
\author[a,b]{Vincenzo Greco}

\affiliation[a]{Laboratori Nazionali del Sud, INFN-LNS,\\
  Via S. Sofia 62, I-95123 Catania, Italy}

\affiliation[b]{Department of Physics and Astronomy, University of Catania,\\
Via S. Sofia 64, 1-95125 Catania, Italy}

\emailAdd{sunyfphy@lns.infn.it}
\emailAdd{salvatore.plumari@ct.infn.it}
\emailAdd{greco@lns.infn.it}

\abstract{Strong electromagnetic fields are expected to be generated in off-central relativistic heavy ion collisions, which can induce a splitting of the directed flow of charged particles and anti-particles ($\Delta v_1$). Such a splitting manifests even for neutral charmed mesons pairs ($D^0,\overline{D}^0$), hence being a direct probe of the formation of deconfined phase with charm quarks as degree of freedom.

In the limit of large $p_T$ and weak interaction with the QGP, a formula of $\Delta v_1(p_T,y_z)$ of charged particles and anti-particles as a function of $p_T$ and rapidity $y_z$ can be obtained, which is found to be related to the spectra of charged particles and the integrated effect of the Lorentz force. This formula is expected to be valid to heavy quarks and leptons at high $p_T$, where the modification to their equations of motion due to the interaction with both QGP and electromagnetic fields is small, and should have a general application. We also proposed a measurement  of $\Delta v_1(p_T,y_z)$ of leptons from $Z^0$ decay and its correlation to that of $D$ mesons, which would be a strong probe determining whether the large splitting measured in experiments has the electromagnetic origin.
}

\FullConference{%
  HardProbes2020\\
  1-6 June 2020\\
  Austin, Texas}

\begin{document}
\maketitle

\section{Introduction}
A very strong magnetic field is generated in  off-central relativistic heavy ion collisions, which triggers in the last decades intense studies of physics related to the parity and charge-parity symmetry breaking process of QCD~\cite{Kharzeev:2007jp,Fukushima:2008xe}. On the other hand, there are other efforts  concentrating on the direct probe of electromagnetic fields, where the directed flow relative to the reaction plane is one of the promising probes~\cite{Gursoy:2014aka,Das:2016cwd,Gursoy:2018yai,Chatterjee:2018lsx}. Besides these numerical studies, we tried to find a general signature of the directed flow induced by e.m. fields in our previous work~\cite{Sun:2020wkg}, and in this proceeding we will specify more details of this work and try to convey the robustness of our conclusion. The signature should be applicable to heavy quarks and leptons, where the interaction with quark-gluon plasma (QGP) is not strong enough to dirty our conclusion.

\section{Collective flows induced by electromagnetic fields in the limit of large $p_T$ and no interaction with QGP}
We adopt the common convention on the configuration of off-central heavy ion collisions, with the center of nuclei moving in positive $z$ direction located in positive $x$ axis, which generates an electromagnetic field with a huge magnetic component along  the negative $y$ direction. In the deconfined phase, charged quarks can interact with such electromagnetic field, but also with the QGP. Though the interaction between the charged quarks with the QGP is strong, it should be much weaker for heavy quarks ($c, b$) compared to light quarks ($u, d, s$)~\cite{Dong:2019unq}, and it becomes further weaken for high $p_T$ heavy quarks as well as leptons of all $p_T$. To derive a compacted formula of the directed flow $v_1$ induced by electromagnetic fields, we thus first neglect the effect of the interaction with QGP, and the modification of it will be discussed in the next section.

In general, electromagnetic fields move charged particles from initial momentum $\mathbf{p}_i$ to the final one $\mathbf{p}_f=\mathbf{p}_i+\mathbf{\Delta}$, where $\mathbf{\Delta}$ is a distribution of the shift in the momenta of particles. In the evaluation of collective flows of charged particles relative to the reaction plane, the relevance is the shift in transverse momentum, where we can use a  transition function $T(\Delta_x,\Delta_y,p_x,p_y,y_z)$ to represent all effects of e.m. fields that shift one particle's initial $p_x$ and $p_y$ at rapidity $y_z$. Thus the distribution function of charged particles after the effect due to e.m. fields becomes:
\begin{eqnarray}
\frac{dN}{dp_xdp_ydy_z}&=&\int d\Delta_x d\Delta_y dp_x^{i} dp_y^{i} \delta(p_x-p_x^i-\Delta_x) \delta(p_y-p_y^i-\Delta_y) f(p_x^i,p_y^i,y_z)T(\Delta_x,\Delta_y,p_x^i,p_y^i,y_z)\nonumber
\\&=&\int d\Delta_x d\Delta_y f(p_x-\Delta_x,p_y-\Delta_y,y_z)T(\Delta_x,\Delta_y,p_x-\Delta_x,p_y-\Delta_y,y_z)\nonumber
\\&\approx&\int d\Delta_x d\Delta_y [f(p_x,p_y,y_z)T(\Delta_x,\Delta_y,p_x,p_y,y_z)-\frac{\partial fT}{\partial p_x} \Delta_x-\frac{\partial fT}{\partial p_y} \Delta_y]\nonumber
\\&=& f-(\frac{\partial f\overline{\Delta}_x}{\partial p_x}+\frac{\partial f\overline{\Delta}_y}{\partial p_y}),
\label{distribution}
\end{eqnarray}
where $f$ is the initial spectra of charged particles and depends only on $p_T$ in boost-invariant approximation, and $\overline{\Delta}_x(p_x,p_y,y_z)$ and $\overline{\Delta}_y(p_x,p_y,y_z)$ are the average shifts in the transverse momenta of a particle with initial $p_x, p_y$ and $y_z$. The approximation in Eq. (\ref{distribution}) is valid when the shifts (much smaller than 1 GeV in most configurations of e.m. fields) are small compared to initial momenta. 

One can adopt a similar idea in the study of collective flows and express $\overline{\Delta}_x$ and $\overline{\Delta}_y$ as a function of azimuthal angle $\phi$ in momentum space. Because the colliding system is symmetric with $y\leftrightarrow-y$, in momentum space one can find  $\overline{\Delta}_x(p_T,\phi,y_z)=\overline{\Delta}_x(p_T,2\pi-\phi,y_z)$ and $-\overline{\Delta}_y(p_T,\phi,y_z)=\overline{\Delta}_y(p_T,2\pi-\phi,y_z)$, and so we have:
\begin{eqnarray}
\overline{\Delta}_x(p_T,\phi,y_z)&=&\sum 2\rm{a_n}(p_T,y_z) \rm{cos}(n\phi), 
\\\overline{\Delta}_y(p_T,\phi,y_z)&=&\sum 2\rm{b_n}(p_T,y_z) \rm{sin}(n\phi).
\end{eqnarray}

Because $B_z$ is absent even in the conducting medium~\cite{Tuchin:2013apa,Li:2016tel} due to colliding nuclei moving only along $z$ directions, one can have $\Delta_x=\int dt q(E_x-v_zB_y)$ and $\Delta_y=\int dt q(E_y+v_zB_x)$ with $v_z=\rm{tanh}y_z$. So $a_n$ and $b_n$ depend weakly on $p_T$, and the azimuthal angle distributions of $\overline{\Delta}_x$ and $\overline{\Delta}_y$ in momentum space measure directly the azimuthal angle distributions of $\bf{E}$ and $\bf{B}$ in coordinate space. Eq. (\ref{distribution}) can thus be further simplified to:
\begin{eqnarray}
f^{'}&=& f-(\rm{}\frac{\partial f\overline{\Delta}_x}{\partial p_T} cos(\phi)-\frac{\partial f\overline{\Delta}_x}{\partial \phi} \frac{sin(\phi)}{p_T}+\frac{\partial f\overline{\Delta}_y}{\partial p_T}sin(\phi)+\frac{\partial f\overline{\Delta}_y}{\partial \phi} \frac{cos(\phi)}{p_T})\nonumber
\\&=&f-\rm{} [\frac{a_1+b_1}{p_T}f+\frac{\partial f}{\partial p_T}(a_1+b_1)]-\rm{} \frac{\partial f}{\partial p_T}(a_0+b_0)cos\phi \nonumber
\\&-&\rm{}\sum_{n=1} [\frac{\partial f}{\partial p_T}(a_{n+1}+b_{n+1}+a_{n-1}-b_{n-1}) +\frac{(n+1)(a_{n+1}+b_{n+1})-(n-1)(a_{n-1}-b_{n-1})}{p_T}f ]cosn\phi ,\nonumber
\\ 
\label{distribution2}      
\end{eqnarray}
where we can find all collective flow coefficients from it.

Eq. (\ref{distribution2}) shows that the distribution is symmetric  with $y\leftrightarrow-y$. On the other hand, in the medium with non-zero chiral magnetic conductivity, $\overline{\Delta}_x$ ($\overline{\Delta}_y$) is not symmetric (anti-symmetric) with $y\leftrightarrow-y$, because $B_y$ ($B_x$) is not symmetric (anti-symmetric) with $y\leftrightarrow-y$ and $B_z$ becomes non-zero~\cite{Li:2016tel,Inghirami:2019mkc}. This means that Eq. (\ref{distribution2}) should have $\rm{sin}(n\phi)$ terms in principle. The disagreement between these two is that the emergence of chiral magnetic conductivity breaks parity ($y\leftrightarrow-y$) symmetry~\cite{Kharzeev:2007jp,Fukushima:2008xe}.

\section{The splitting of the directed flow of charged particles}
Using Eq. (\ref{distribution2}), one can obtain the directed flow in the limit of large $p_T$~\cite{Sun:2020wkg}:
\begin{eqnarray}
v_1(p_T,y_z)=\frac{\int d\phi f^{'} cos\phi}{\int d\phi f^{'}}=-\frac{\frac{\partial f}{\partial p_T}(a_0+\frac{a_2+b_2}{2})+\frac{a_2+b_2}{p_T}f}{f-\frac{a_1+b_1}{p_T}f-\frac{\partial f}{\partial p_T}(a_1+b_1)}\approx-\rm{}\frac{\partial lnf}{\partial p_T}a_0(p_T,y_z),
\label{v1}      
\end{eqnarray}
where the approximation is obtained by keeping the effect of e.m. fields in the leading order and by knowing the smallness of  the quadrupole moments of e.m. fields in the overlapping region of colliding nuclei. 

It seems to be surprising that $v_1$ depends on $\rm{}\frac{\partial lnf}{\partial p_T}$ at first sight. However, one can understand it better by studying charged particles with a spectra uniformly distributed in $p_T$. If Lorentz force moves the momenta of all charged particles along positive $x$ axis with a constant shift, one can easily find that the distribution is exactly same as the initial one after the shift unless one looks at the boundary of the final spectra, which means that $v_1$ is still zero after this shift.

The calculation of $a_0(p_T,y_z)$ is complicated due to the complex time and spatial distributions of $E_x$ and $B_y$. However, we can find a compact formula by assuming $eB_y(x,y,t,z)=-B(\tau)\rho_B(x,y)$ with $\rho_B(x,y)=\rm{exp}[-\frac{x^2}{2\sigma_x^2}-\frac{y^2}{2\sigma_y^2}]$. In this case $a_0(p_T,y_z)$ can be evaluted from our study~\cite{Sun:2020wkg}, and finally $\Delta v_1$ of positively and negatively charged particles becomes:
\begin{eqnarray}
\Delta v_1(p_T,y_z) \propto -\rm{}|q|\frac{\partial lnf}{\partial p_T}\int_0^{y_z}\frac{d\chi}{\rm{cosh}\chi}\left[ \tau_2B(\tau_2)-\tau_1B(\tau_1)\right],
\label{final}
\end{eqnarray}
with $\tau_{1}=\frac{\tau_0\rm{cosh}y_z}{\rm{cosh}\chi}$ and $\tau_{2}=\frac{(\tau_0+Rm_T/p_T)\rm{cosh}y_z}{\rm{cosh}\chi}$. $\tau_{1,2}$ can be treated as charged particle's formation time and the escape time out of the electromagnetic field, and is independent of $p_T$ at $p_T\gg m$. In fact, $a_0(p_T,y_z)$ should be independent of $p_T$ as long as $p_T\gg m$ for any configurations of e.m. fields, because the trajectory of particles and the Lorentz force experienced by them are same due to the fact that the velocity of particles is independent of $p_T$ in this limit.

Though Eq. (\ref{final}) is obtained with a very strong assumption of e.m. fields, it may be robust at small $y_z$, because the azimuthal angle distribution of e.m. fields is not so relevant based on our analysis of the former section, and $E_x$ and $B_y$ vary little in the overlapping region and at small space-time rapidity. Eq. (\ref{final}) thus provides a general scaling of the $d\Delta v_1/dy_z (p_T)$ of charged particles.

In Ref~\cite{Sun:2020wkg}, we studied $\Delta v_1$ of charged leptons $l^+-l^-$ from $Z^0$ decay and $D^0-\overline{D}^0$ using the e.m. field that can reproduce the large $\Delta v_1$ of $D^0-\overline{D}^0$ measured in experiments at 5.02 TeV Pb+Pb collisions~\cite{Acharya:2019ijj}, and found that 
$\Delta v_1(l)$ has a jump around 45 GeV$/c$, and its magnitude is always smaller than $\Delta v_1(D^0)$. It  can be understood by Eq. (\ref{final}) from the peculiar spectra of leptons from the decay of $Z^0$, and this surprising finding can be used to determine whether the measurement of $\Delta v_1$ has the e.m. field origin.
\section{The effect of the interaction with QGP}
Heavy quarks can interact with QGP and the effect of e.m. fields may be weaken by that. We thus adopted the standard Fokker-Planck equation with the interaction strength tuned to successfully reproduce the nuclear modification factor $R_{AA}$ and  elliptic flow $v_2$ of $D$ mesons~\cite{Dong:2019unq,Sun:2019fud} at both RHIC and LHC energies, and studied this effect numerically.

\begin{figure}[h]
\centering
\includegraphics[width=0.5\linewidth]{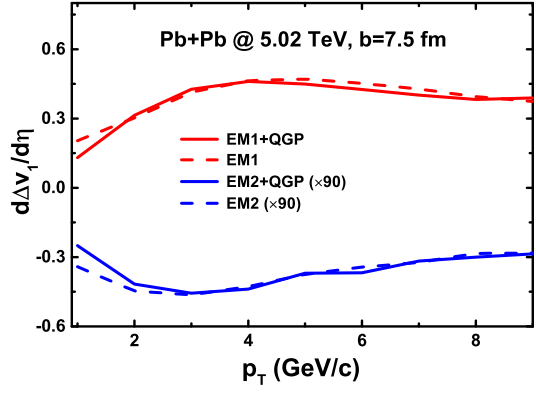}
\caption{(Color online) $p_T$ dependence of $d\Delta v_1^D/d\eta$ of $D^0$ and $\overline{D}^0$  for two configurations of e.m. fields plus with or without the interaction with QGP.
}
\label{fig:dv1pt}
\end{figure}

We have picked up two typical configurations of e.m. fields: EM1 is the e.m. field that can reproduce experimental measurements of $d\Delta v_1^D/d\eta$~\cite{Acharya:2019ijj}; EM2 is the one calculated using a constant medium conductivity in the lower limit of LQCD calculations~\cite{Gursoy:2014aka,Das:2016cwd,Gursoy:2018yai,Chatterjee:2018lsx} that is adopted by a lot of studies but with a discontinuity in the time evolution of e.m. fields. Case EM1  leads to a positive  $d\Delta v_1^D/d\eta$ with its magnitude order two larger than case EM2, because the difference between $\tau B$ at the freeze-out time and the formation time of charm quarks is positive and larger for EM1. For the effect of the interaction with QGP, it is seen in Fig. \ref{fig:dv1pt} that the interaction  damps $d\Delta v_1^D/d\eta$ only at $p_T<2$ GeV, and the suppression is negligible at high $p_T$ for these two significantly different e.m. fields. It is also seen in Fig. \ref{fig:dv1pt} that at high $p_T$, $d\Delta v_1^D/d\eta$ shares a similar $p_T$ dependence for these two e.m. fields even though the huge gap between their magnitude. This is because the variation of $d\Delta v_1^D/d\eta$ with $p_T$ mostly  comes from $\rm{}\frac{\partial lnf}{\partial p_T}$ (see Eq. (\ref{final})).

\section{Conclusions}
We have obtained a compact formula for the $\Delta v_1$ of charged particles induced by electromagnetic fields, which is shown to be applied to heavy quarks and leptons at high $p_T$. According to our step by step analysis, we think this formula is robust and can be used to test whether the experimental measurements of $\Delta v_1$ have an e.m. origin.

\end{document}